\documentclass[aps,prd,twocolumn,nofootinbib,superscriptaddress]{revtex4-2}
\usepackage{graphicx} % Required for inserting images

\usepackage{amsmath,amssymb,bm}
\usepackage{graphicx}
\usepackage{hyperref}
\usepackage[capitalise]{cleveref}
\usepackage{xcolor}
\usepackage{booktabs}

\newcommand{\oja}{\textsc{OmniJet}-$\alpha$}

\begin{document}

\title{Neural Scaling Laws for Jet Generation}

\author{Oz Amram}
\affiliation{Fermi National Accelerator Laboratory, Batavia, IL 60510, USA}

\author{Darius A. Faroughy}
\affiliation{NHETC, Dept.\ of Physics and Astronomy, Rutgers University, Piscataway, NJ 08854, USA}

\author{Tjarko Gerdes}
\affiliation{Institute for Experimental Physics, Universität Hamburg,
	Luruper Chaussee 149, 22761 Hamburg, Germany}

\author{Anna Hallin}
\affiliation{Institute for Experimental Physics, Universität Hamburg,
	Luruper Chaussee 149, 22761 Hamburg, Germany}

\author{Gregor Kasieczka}
\affiliation{Institute for Experimental Physics, Universität Hamburg,
	Luruper Chaussee 149, 22761 Hamburg, Germany}

\author{Michael Kr\"amer}
\affiliation{Institute for Theoretical Particle Physics and Cosmology, RWTH Aachen University, Aachen, Germany}

\author{Humberto Reyes-Gonzalez}
\affiliation{Institute for Theoretical Particle Physics and Cosmology, RWTH Aachen University, Aachen, Germany}

\author{David Shih}
\affiliation{NHETC, Dept.\ of Physics and Astronomy, Rutgers University, Piscataway, NJ 08854, USA}

\date{\today}

\begin{abstract}
Recently observed empirical scaling laws describe the performance of foundation-type models as three independent key quantities --- dataset size, compute, and model parameters --- are modified.
Extracting these scaling laws informs the training of large complex models for which the tuning of hyperparameters in traditional ways is not feasible.
This work for the first time explores if scaling laws can also be observed for the task of particle jet generation --- both relevant as a pre-training objective for foundation models and as in-situ simulation by itself. 
We indeed replicate the key logarithmic scaling law behavior for model-size scaling. Beyond studying the next token prediction validation loss of the generative model, we also study the sliced Wasserstein distance of five physical quantities that are not immediately available to the model during training. Our study shows that this quantity is monotonically related to the next token prediction validation loss, meaning that this loss is indeed a good proxy for the physics performance. For the scaling with dataset size and compute, we observe substantially weaker scaling behavior of both the loss and the sliced Wasserstein distance. We analyze this behavior by introducing the concept of a learnable window, and argue that autoregressive next token prediction on jet constituents exhibits comparatively rapid saturation relative to language-model studies. We discuss possible origins of this behavior, including the stochastic nature of QCD radiation and differences between generative and supervised learning tasks in collider physics.

\end{abstract}

\maketitle

\section{Introduction}

Neural scaling laws have emerged as one of the central empirical discoveries underlying modern artificial intelligence~\cite{kaplan2020scalinglawsneurallanguage}. The validation losses in a broad range of deep learning systems --- most prominently large language models~\cite{brown2020language} --- exhibit remarkably robust power-law scaling behavior as a function of model size, dataset size, and training compute, over many orders of magnitude. Such scaling laws are of immediate practical importance: they allow frontier-scale training runs to be forecast and optimized before incurring the full computational cost. At the same time, they raise fundamental theoretical questions about the structure of learning itself and the nature of the underlying data distributions.

Despite the intense recent interest in scaling laws, many open questions remain and are being actively researched: how universal these scaling behaviors truly are, how strongly they depend on architecture and optimization choices, or how they are modified by properties of the underlying data. In addition, while scaling exponents are commonly measured using training or validation losses, it is less clear how reliably these quantities translate into downstream, domain-specific performance metrics ~\cite{lourie2025scalinglawsunreliabledownstream}. 

Scaling laws are only beginning to be explored in high energy physics. Recent studies have investigated scaling behavior for collider classification tasks, including top tagging~\cite{batson2023scalinglawsjetclassification}, boosted jet tagging~\cite{Vigl:2026ppx} and heavy flavor identification~\cite{ATLAS:2026zdb}, as well as for surrogate models of scattering amplitudes~\cite{Bahl:2026jvt}. Initial hints of scaling were also observed for the jet generative modeling task in \cite{Amram:2024fjg}.
These studies suggest that at least some of the empirical phenomena observed in mainstream machine learning may also appear in particle physics applications. At the same time, collider data differs qualitatively from natural language and vision data: it is both very stochastic and yet highly constrained by the dynamics of  quantum field theory. It is thus nontrivial and far from obvious that the same scaling behavior seen for natural images and large language models should persist in the HEP domain.

In this work, we systematically investigate neural scaling laws for jet generative models for the first time. We study scaling behavior in the foundation model OmniJet-$\alpha$~\cite{Birk:2025fbs}, an autoregressive GPT-style transformer trained on tokenized jet constituents using next token prediction (NTP). The setup is intentionally close in spirit to the original language-model scaling law studies. This allows us to directly probe whether scaling-law phenomena similar to those observed in large language models also emerge in collider jet generation.

A second distinctive aspect of this work is that we study scaling directly on real collider data rather than on Monte Carlo simulation. We use the Aspen Open Jets (AOJ) dataset~\cite{amram_oz_2024_16505}, derived from CMS Open Data, containing approximately 180 million jets reconstructed from proton-proton collision events. To our knowledge, this is the first investigation of neural scaling laws on experimentally recorded collider data.

We investigate scaling behavior as a function of three quantities: model size $N$, dataset size $D$, and training compute $C$. For each scaling direction, we analyze both the validation NTP loss and physics-motivated metrics of generation quality based on sliced Wasserstein distances (SWD) computed on high-level jet observables. This allows us to study not only whether scaling laws are present in the training objective itself, but also whether improvements in likelihood translate into improvements in physically meaningful observables.

We find clear evidence for scaling-law behavior in the dependence of validation loss on model size. Over more than three orders of magnitude in parameter count, the NTP loss is well described by a power law. Our fit also allows for the extraction of $L_\infty$, the asymptotic floor of the loss. 
Moreover, improvements in the NTP loss correlate strongly with improvements in SWD-based physics metrics, indicating that the observed scaling behavior is physically meaningful and not merely an artifact of the autoregressive training objective. The corresponding SWD values also exhibit approximate power-law behavior with model size and approach the intrinsic statistical floor associated with finite-sample comparisons of real data.

In contrast, scaling with dataset size is substantially less conclusive. Although the data are compatible with a power-law interpretation, the observed dynamic range is small and statistical uncertainties are comparatively large. Several factors may contribute to this weaker signal. Due to computational constraints, only a limited number of independent trainings could be performed for each dataset size, especially at large $D$. In addition, the AOJ dataset is organized into chunks whose statistical composition may fluctuate, potentially introducing nontrivial variance between dataset realizations. More fundamentally, it is possible that the dominant structures in the jet distribution are already learned with relatively modest dataset sizes, leaving only a small reducible component accessible through further scaling.

We also investigate scaling with compute using an isoFLOP analysis analogous to that employed in large language model studies~\cite{hoffmann2022trainingcomputeoptimallargelanguage}. Here we encounter an additional challenge: the models converge extremely rapidly, and even comparatively small models already achieve losses close to the apparent asymptotic limit. As a result, the expected U-shaped isoFLOP curves are unusually flat, making the extraction of compute-optimal model sizes highly uncertain. To characterize this phenomenon, we introduce the concept of a \emph{learnable window}: the gap between the loss of a uniform predictor and the apparent irreducible entropy floor of the dataset. We find that this learnable window is dramatically smaller for jet next-token prediction than for language modeling. In practice, even relatively small models already capture most of the learnable structure in the dataset. This provides a possible explanation for the weak scaling observed with dataset size and compute.

More broadly, our results suggest that scaling behavior in collider physics may depend sensitively on the structure of the underlying task and data distribution. In supervised jet classification tasks, recent work has observed strong continued scaling behavior over large ranges in model and dataset size. By contrast, autoregressive generative modeling of generic QCD jets appears to approach saturation much earlier. This difference may reflect the intrinsically stochastic nature of QCD radiation, the dominance of relatively featureless QCD jets in the AOJ dataset, or the mismatch between autoregressive sequence ordering and the causal structure of parton showers. Understanding these possibilities may provide useful insight not only into machine learning for particle physics, but also into the broader question of when and why neural scaling laws emerge.

The outline of our paper is as follows. In Sec.~\ref{sec:method} the method is explained, which includes an introduction of the model, the dataset, the metrics, the details of the three phases of the study (corresponding to scaling with $N, D$ and $C$), as well as the concept of the learnable window. Sec.~\ref{sec:results} presents the results, followed by a discussion in Sec.~\ref{sec:discussion} and conclusions in Sec.~\ref{sec:conclusion}.

\section{Method}
\label{sec:method}
We study scaling as a function of three key parameters: the model size $N$, the dataset size $D$ and the computational cost $C$. The study is conducted in three phases. Phase 1 investigates the scaling of $N$ given a fixed dataset size and training budget, phase 2 the scaling of $D$ at a fixed model architecture, and phase 3 investigates the scaling of $C$ and addresses the question of optimizing the model size for a given amount of compute. Phase 1 and 2 follow~\cite{kaplan2020scalinglawsneurallanguage}, whereas phase 3 adopts the isoFLOP methodology of~\cite{hoffmann2022trainingcomputeoptimallargelanguage}, training with different combinations of model and dataset sizes for given compute budgets.

\subsection{Model}
\label{sec:model}
\oja~\cite{Birk:2024knn} is a foundation model for jet physics, using next token prediction as training target\footnote{A newer version~\cite{Birk:2025fbs} allows pretraining also on masked token prediction. This work, however, only uses the NTP version.}. It is thus architecturally close to modern large language models. While \oja{} is capable of classification as well as generation, only the generative performance is evaluated in this work. As the name suggests, the model is developed for jet physics. Jets are collimated sprays of particles that are important objects in collider physics. The particles in a jet are referred to as jet constituents, and come with a number of features such as kinematic information, particle ID, etc. A VQ-VAE~\cite{oord2018neural,bao2022beit,huh2023straightening} is used to convert the physics information of the constituents into integer tokens, turning each jet into a sequence of tokens. The set of tokens that the VQ-VAE has access to is called the codebook, and the number of tokens is referred to as the codebook (or vocabulary) size. In contrast to language where the order of words matter, jets are point clouds and the constituents have no meaningful order. They are ordered by transverse momentum $p_\mathrm{T}$ by convention, but this ordering is arbitrary. Once the input data is tokenized, it passes through a GPT~\cite{Radford2018ImprovingLU} style transformer decoder with a causal mask, training on a next token prediction target using cross-entropy loss. It thus functions as a multi-class classifier. In the context of the model parameter counting used in this study, it is important to note that the model contains one embedding layer (from codebook to embedding dimension) and one unembedding layer (from embedding dimension back to codebook). Once trained, the model is able to generate new jets from the learned distribution. The generated sequences can then be decoded back to physics space using the frozen VQ-VAE. 

\subsection{Dataset}
\label{sec:dataset}
The dataset used for this study is the Aspen Open Jets dataset~\cite{amram_oz_2024_16505}, which contains roughly 180 million jets derived from CMS Open Data~\cite{CMS:2016G,CMS:2016H}. The dataset has been divided into smaller files, in total 2,345, with an average of 70k jets per file. As this is real observed data, the jets do not come with any class or type labels. The majority of the jets are however expected to originate from light quarks and gluons, referred to as ``QCD jets'', while an estimated $\mathcal{O}(10^5)$ jets come from vector bosons or top quarks.

Only the kinematic features of the constituents are used in the study: transverse momentum  ($p_\mathrm{T}$), and relative angle of the constituent with respect to the jet axis ($\Delta \eta$ and $\Delta \phi$). The average number of constituents per jet is 49. For each jet, up to the top 128 highest-$p_T$ constituents are included, with any remaining slots padded to maintain a fixed sequence length. A VQ-VAE with a codebook size of 32,768 is used to tokenize the jets.

\subsection{Metrics}
\label{sec:metrics}
The scaling laws for two metrics are analyzed. The first is the next token prediction (NTP) validation loss, averaged first over the number of non-padding tokens in each jet and then over all 500 validation batches, similarly to \cite{kaplan2020scalinglawsneurallanguage,hoffmann2022trainingcomputeoptimallargelanguage}. The validation loss measures the information-theoretic convergence of the autoregressive model through the expected log-likelihood and is expected to asymptotically approach the intrinsic entropy of the data distribution (see App.~\ref{app:loss}). The second is the sliced Wasserstein distance (SWD)~\cite{Bonneel2015} based on five high-level jet observables. These are the transverse momentum of the jet $p_\mathrm{T}$, the jet mass $m$, the n-subjettiness ratios~\cite{Thaler:2010tr} $\tau_{21}$ and $\tau_{32}$, and the number of jet constituents$n$. SWD thus probes the generative performance for the jet as a whole, based on variables that were not immediately available to the model during training.  The SWD is a frequentist goodness-of-fit metric computed directly on empirical distributions, quantifying how accurately the generated samples reproduce the feature space of the selected physical observables. The SWD is calculated for each model by comparing 50k generated jets with 50k real jets unused in model training. Further details on the SWD implementation can be found in App.~\ref{app:swd}. 

The power laws for the NTP loss as a function of the studied variable $X$ are of the form
\begin{equation}
    L(X) = A_X X^{-\beta_X} + L_{\infty},
\end{equation}
where $A_X$ is an constant, $\beta_X$ is the scaling exponent, and $L_{\infty}$ is the irreducible loss, which corresponds to the entropy of the true distribution (see App.~\ref{app:loss}). 

The SWD power law has a similar form as the one for the loss,
\begin{equation}
    \mathrm{SWD}(X) = A_X^\mathrm{SWD} X^{-\beta_X^\mathrm{SWD}} + \mathrm{SWD}_f,
\end{equation}
where $\mathrm{SWD}_f$ is the \textit{SWD floor}. It is measured by comparing two sets of 50k jets drawn from the distribution of real jets after having been tokenized and reconstructed \footnote{A separate SWD floor calculation using the original (non-tokenized) resolution results in the same value (within error bars). Thus, the tokenized resolution does not inflate $\mathrm{SWD}_f$.}. $\mathrm{SWD}_f$ thus corresponds to the inherent variability of the chosen features between the two samples given their limited size (50k). In principle, a more powerful metric and/or a larger evaluation set would exhibit a different floor (see \cite{Grossi:2024axb,Grossi:2025pmm,Cappelli:2025myc}). The uncertainty on the SWD values in this work is estimated via 50 bootstrap resamples of 50k jets drawn with replacement, and the quoted mean and standard deviation are those of the bootstrapping.  

Thus, the two metrics probe different aspects of scaling. The scaling of the NTP loss probes the improving performance on the direct training task, while the SWD metric probes how these gains translate to improved modeling of high-level emergent observables. Studying both metrics allows us to characterize how improvements in likelihood translate into physically meaningful gains in sample quality.

\subsection{Phases of the study}
The study is carried out in three phaces, corresponding to the three variables $N, D$ and $C$. The search space for all three phases is shown in Figs.~\ref{fig:methodology_ND_space} and \ref{fig:methodology_NC_space} in App.~\ref{app:methodology}.

\subsubsection{Phase 1: model size}
Phase 1 studies performance as a function of model size. 
Nine model sizes are used, spanning 3.5 orders of magnitude in non-embedding parameters $N_{\mathrm{non-emb}}$, from Pico ($N=2.5\times 10^4$) to XXL ($N=8.5\times 10^7$). The model size is varied by changing the embedding dimension, the number of attention heads, the number of GPT blocks and the head dimension. Further details can be found in App.~\ref{app:search_space}.

Since the only bottleneck in phase 1 is supposed to be $N$, all trainings have access to the full dataset and a compute budget of 400,000 gradient update steps. The compute budget was determined after preliminary studies with a constant learning rate showed that all models converged before this point. The actual trainings then used a cosine annealing schedule with the minimum learning rate set to zero. This schedule forces the loss to plateau during the cooldown. Given that we are not in an overfitting regime, the best validation loss will by construction occur in this plateau region. Since the loss varies very little in this region it does not matter that much which exact checkpoint is chosen, and so for simplicity the last checkpoint of the training is used. The per-point uncertainty is estimated as the standard deviation of the last 20 validation cycles (corresponding to steps 362k-400k). Confidence intervals on the fit parameters for the loss is obtained via 2000 parametric bootstrap resamples. For the SWD fit, the parameter uncertainties are the $1\sigma$ standard errors from the non-linear least-squares fit.

\subsubsection{Phase 2: dataset size}
Phase 2 examines the data efficiency of the model: all else being equal, how does the amount of data affect the performance? Varying the amount of data while removing any bottleneck from compute, leads to multiple passes over the scheduled data for the smallest dataset sizes. At the lowest $D$ ($6.4\times 10^6$), the model has 196 scheduled passes, while at the highest ($8.1\times 10^9$) the corresponding number is 0.63. Model L is chosen for the phase 2 study, as it is large enough to not bottleneck the measurement of the loss $L(D)$. The results of phase 1 in Sec.~\ref{sec:results_phase_1} show that the two larger models only provide a modest increase in performance. Model L is thus a good choice, as it performs well enough while still keeping the required compute resources manageable. 

Eight dataset sizes were chosen, spanning three orders of magnitude in $D$. The full list can be found in App.~\ref{app:search_space}. For small dataset sizes, the trainings were repeated three times with different seeds for the file selection. For the larger dataset sizes, any potential file selection effects that could affect the actual physics content of the data would be diluted, which is why the runs for these were not repeated.

The training itself was performed the same way as for phase 1, with 400k gradient update steps and cosine annealing of the learning rate. For the smallest $D$, this standard training loop would require more jets per epoch than are available. The compute budget was therefore lowered to 100k gradient updates, which from the observed overfitting was more than enough. 

Consistent with~\cite{kaplan2020scalinglawsneurallanguage}, the checkpoint for evaluation is the checkpoint with lowest validation loss (rather than the last checkpoint as for phase 1). In the power law fit, the irreducible loss $L_{\infty}$ is clamped to $\min L(D)$, as the fit would otherwise place $L_{\infty}$ above the loss for the largest $D$. Per-point error estimates are taken from the variance across seeds. Single seed runs do not get error estimates.  Confidence intervals on the fit parameters for the loss is obtained via 5000 parametric bootstrap resamples. For the SWD fit, the parameter uncertainties are the $1\sigma$ standard errors from the non-linear least-squares fit.

\subsubsection{Phase 3: compute}
Phase 3 studies scaling with compute, following the isoFLOP approach of~\cite{hoffmann2022trainingcomputeoptimallargelanguage}. We use five compute budgets, spanning one order of magnitude in FLOPs and spaced approximately uniformly in $\log C$. The connection between compute budget, model size and gradient steps is
\begin{equation}
    C = 6 N_{\mathrm{incl}} B S_{\mathrm{seq}} n_{\mathrm{steps}},
\label{eq:compute}
\end{equation}
where $C$ is the compute budget in FLOPs, $N_{\mathrm{incl}}$ is the number of model parameters excluding the embedding layer but including the unembedding layer~\cite{porian2024resolving}, $S_{\mathrm{seq}}$ is the sequence length and $n_{\mathrm{steps}}$ is the number of gradient steps. The factor of 6 is adopted from~\cite{kaplan2020scalinglawsneurallanguage}, and accounts for the forward and backward passes. The full list with the corresponding number of gradient update steps for each model can be found in Tab.~\ref{tab:isoflops} in App.~\ref{app:search_space}. A smaller model, Sub-Pico, was added for phase 3, due to the observed flatness of the isoFLOP curves. Some $N$ -- $C$ combinations are not viable in this context: runs are excluded if 
\begin{itemize}
    \item the number of jets expected to be seen exceeds the number of unique training jets, or
    \item the number of training steps is below 2500.
\end{itemize}
A small number of additional $N$ -- $C$ combinations were omitted for practical reasons: very small models that at high compute budgets would require prohibitive wall-clock time, and $N$ -- $C$ combinations with near overlap with runs from phase 1.

The training was performed with a cosine annealing schedule with the minimum learning rate set to zero. The number of steps is decided by the compute budget and model size according to Eq.~\eqref{eq:compute}, and ranges from 5k to 475k. All runs use the full training dataset of $D=1\times 10^9$ tokens. Two runs (XS at $C=5\times 10^{17}$ and M$^-$ at $C=10^{18}$) leads to mild data repetition (a factor of 1.17 and 1.04, respectively). Using the $L(D)$ power law from Sec.~\ref{sec:results_phase_2}, the estimated loss penalty from this data repetition is below the per-run noise.

The optimal model size for the given compute budget, $N^*(C)$, is determined by parabolic interpolation of the isoFLOP curves rather than discrete selection. Uncertainty is estimated using 2,000 bootstrap resamples (with replacement), refitting the parabola for each, and extracting the resulting $N^*(C)$. The power law for $N^*(C)$,
\begin{equation}
    N^*(C) \propto C^a
\end{equation}
is fitted using the $N^*$ value for each of the five compute budgets $C$. Additionally, the loss achieved by the compute-optimal model per budget can be extracted and fitted to a power law
\begin{equation}
    L^*(C) = A_C C^{\beta_C}+L_{\infty}.
\end{equation}
Since we only have five budgets, this will be reported as a point estimate.

\subsection{Learnable window}
The loss $L_{\mathrm{NTP}}$, measured in all three phases, cannot be directly compared to the corresponding loss from the language domain. However, comparing the asymptotic behavior of NTP loss across different domains can reveal insights about the underlying data structures. The cross-entropy loss for the next token prediction task for a data distribution $p$ and a model $q$ can be written as
\begin{equation}
    H(p,q) = H(p) + D_{\mathrm{KL}}(p||q).
\label{eq:cross-entropy}
\end{equation}
This loss is bounded from below by the entropy $H(p)$ of the next token given its context under $p$.  
The worst a non-pathological model could do is to assign equal probabilities to all tokens in the codebook of size $V$: $q_{\mathrm{uniform}}(x)=1/V$. The leads to an upper bound loss from such a predictor of $L_{\mathrm{uniform}}=\log V$. 

As a metric to compare how structured datasets are across different domains, we introduce the \textit{learnable window} $W$, defined as the gap between these two bounds,
\begin{equation}
    W = \log V - H(p).
\end{equation}
We note that the learnable window is specific to the data type (which provides $H(p)$) and tokenization setup (which determines $V$). In this work, we use the $L_{\infty}$ from phase 1 as an estimate for $H(p)$, as it is the capacity-independent term of the power law \eqref{eq:L_N} as well as of the cross-entropy loss \eqref{eq:cross-entropy}. 
A perfectly deterministic dataset, in which the next token in the sequence could be perfectly predicted from the context, would have $H(p) = 0$ and maximal learnable window $W=\log V$. 
Whereas an entirely stochastic dataset, in which the context provides no information in determining the next token, would have a very small learnable window.

We will analyze three metrics:
\begin{enumerate}
    \item \textbf{Window exploitation} $\Delta L/W$ measures how much of the learnable window a scaling step traverses.
    \item \textbf{Irreducibility ratio} $H(p)/\log V$ is the fraction of the total loss range that is occupied by the irreducible loss.
    \item \textbf{Effective perplexity} $\mathrm{PPL}= e^{H(\hat{p},q)}$ measures how perplexed the model that assigns probabilities $q$ was on the test set with empirical probabilities $\hat{p}$. This can be interpreted as the number of equally likely next-token options the model faces. We will use $L_{\infty}$ as an estimate for $H(\hat{p},q)$, giving the perplexity as $\mathrm{PPL}= e^{L_{\infty}}$.
\end{enumerate}
Note that the first two of these metrics directly depend on $V$, while the third has an indirect dependence in that a larger codebook size makes the choices more difficult for the model. This means that if we want to compare our results to the language domain, we should find a model that has a similar codebook size. Our codebook size of 32,768 is similar to that of Ref.~\cite{hoffmann2022trainingcomputeoptimallargelanguage}.

\section{Results}
\label{sec:results}
\subsection{Phase 1: model size}
\label{sec:results_phase_1}
\begin{figure}
    \centering
    \includegraphics[width=0.95\linewidth]{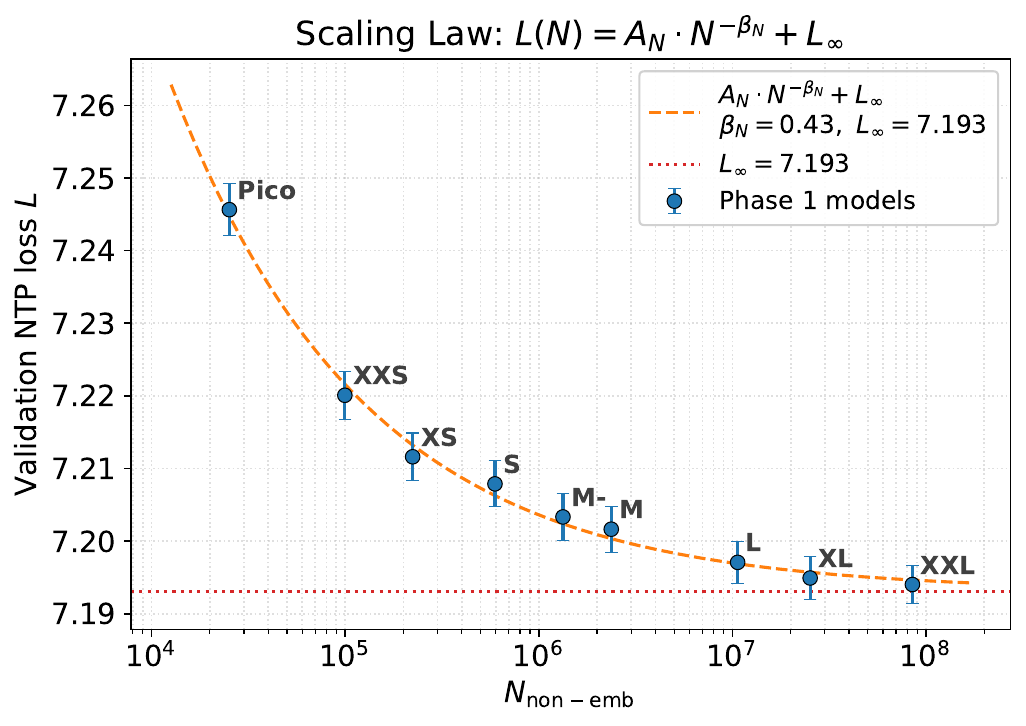}
    \caption{The phase 1 NTP loss $N_\mathrm{non-emb}$ can be fitted to a power law as a function of the non-embedding parameters. Models are evaluated at the last checkpoint, and the per-point error bars are estimated as the standard deviation of the last 20 validation cycles. The orange dashed curve shows the fitted scaling law, and the red dotted line shows the fitted $L_{\infty}$.}
    \label{fig:Lntp}
\end{figure}

Figure \ref{fig:Lntp} shows the last-checkpoint NTP validation loss for the nine models as a function of the non-embedding parameters $N$. 
The NTP loss is described by a power law 
\begin{equation}
    L(N) = A_N N^{-\beta_N}+L_{\infty}
\label{eq:L_N}
\end{equation}
where 
\begin{equation*}
    A_N=4.15, \qquad \beta_N = 0.43, \qquad L_{\infty} = 7.193~\mathrm{nats}
\end{equation*}
with 95\% confidence intervals $[0.94,30.54]$ for $A_N$, $[0.28,0.63]$ for $\beta_N$ and $[7.186, 7.198]$~nats for $L_{\infty}$.

\begin{figure}
    \centering
    \includegraphics[width=0.95\linewidth]{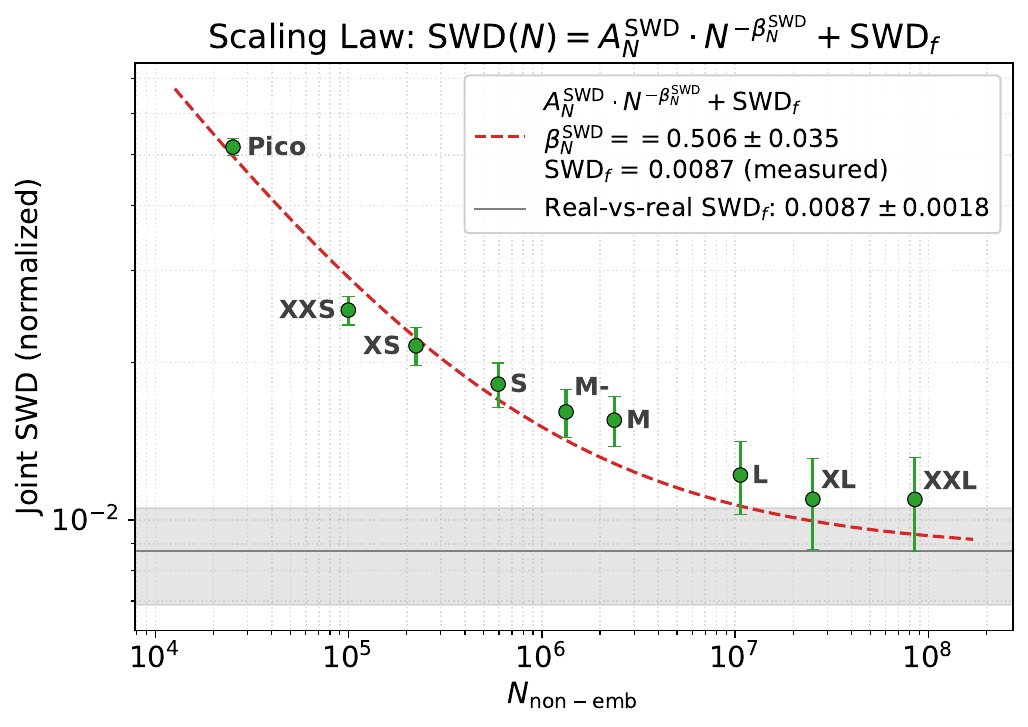}
    \caption{The SWD of phase 1 can be fitted to a power law as a function of the number of non-embedding model parameters $N_{non-emb}$. The per-point uncertainties are estimated via bootstrapping as described in the text. The power law is indicated by the red dashed curve. The real-vs-real SWD floor is shown as a grey line with a corresponding uncertainty band.}
    \label{fig:SWDntp}
\end{figure}

The corresponding SWD is shown in figure \ref{fig:SWDntp}, with the measured real-vs-real noise floor SWD$_f=0.0087\pm 0.0019$. The corresponding scaling law is
\begin{equation}
    \mathrm{SWD}(N) = A_N^{\mathrm{SWD}} N^{-\beta_N^{\mathrm{SWD}}}+\mathrm{SWD}_f
\end{equation}
where 
\begin{equation*}
    A_N^{\mathrm{SWD}}=6.9\pm 2.7, \qquad \beta_N^{\mathrm{SWD}} = 0.506\pm 0.035.
\end{equation*}

\begin{figure}
    \centering
    \includegraphics[width=0.95\linewidth]{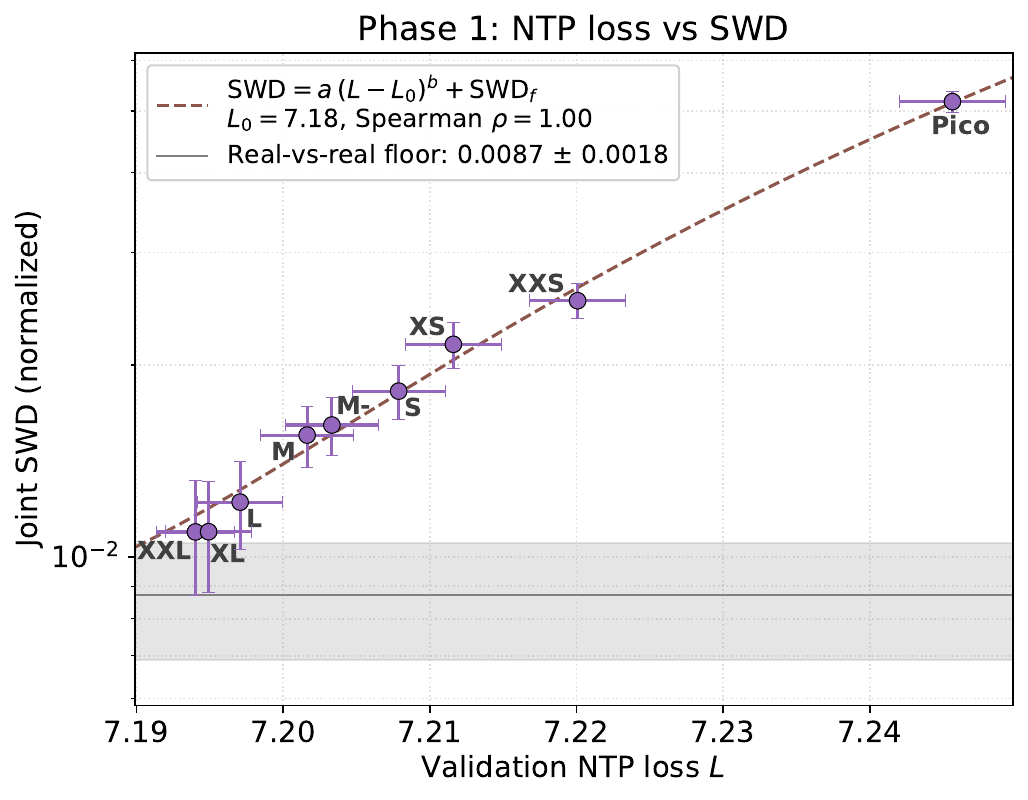}
    \caption{The phase 1 NTP loss and SWD have a monotonic relationship, such that the SWD can be fitted to a function of the NTP loss. The dashed curve shows this fit, and the grey line and corresponding uncertainty band shows the real-vs-real SWD floor.}
    \label{fig:N_L_ntp_vs_swd}
\end{figure}

Figure \ref{fig:N_L_ntp_vs_swd} shows the association between $L(N)$ and $\mathrm{SWD}(N)$. They show a clear monotonic correlation, meaning that in terms of model size $L_{\mathrm{NTP}}$ is a good proxy for the physics performance measured by SWD.

\subsection{Phase 2: dataset size}
\label{sec:results_phase_2}
\begin{figure}
    \centering
    \includegraphics[width=0.95\linewidth]{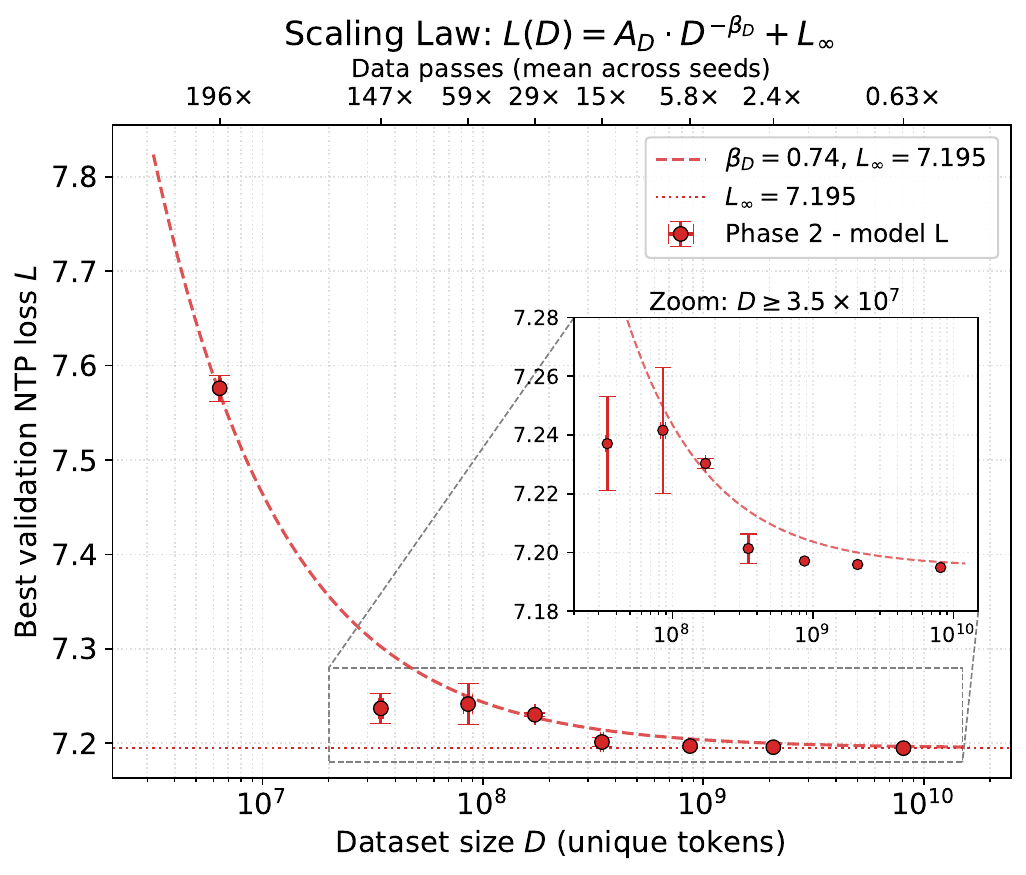}
    \caption{The NTP loss fitted to a scaling law as a function of the dataset size, using model size L and the checkpoint from the epoch with the lowest validation loss. The numbers indicate the number of files used. For $D\leq 3.5\times 10^8$, the trainings were repeated three times with different seeds in order to mitigate any stochastic effects arising from the choice of files. Horizontal error bars (barely visible) show the standard deviation of $D$ across seeds. The vertical per-point uncertainties were also taken from the variance over seeds, where available (the three largest $D$ were single-seed runs).}
    \label{fig:phase2c_ntp_loss}
\end{figure}

The minimum validation NTP loss as a function of the unique training tokens $D$ is shown in Fig. \ref{fig:phase2c_ntp_loss}. The loss can be described by a power law of the form
\begin{equation}
    L(D) = A_D D^{-\beta_D}+L_{\infty}
\label{eq:L_D}
\end{equation}
where 
\begin{equation*}
    A_D=4.3\times 10^{4}, \qquad \beta_D = 0.74, \qquad L_{\infty} = 7.195~\mathrm{nats}
\end{equation*}
with 95\% confidence intervals $[3.4\times 10^{3},9.0\times 10^{4}]$ for $A_D$, $[0.58,0.79]$ for $\beta_D$ and $[7.170,7.195]$~nats for $L_{\infty}$. The value for $L_{\infty}$ obtained here agrees with the one obtained in phase 1.

The power law \eqref{eq:L_D} is dominated by the point $D=6.4\times 10^6$. Removing this point shrinks the fitted dynamic range by a factor of 9. The remaining points are already very close to the saturation regime. It is worth noting that the difference in loss between $D=6.4\times 10^6$ and $D=3.5\times 10^7$ is not a consequence of the \textit{scheduled} data passes, but of the \textit{productive} passes, ie the number of passes over the data that the model has completed before it reaches its minimum validation loss. For $D=3.5\times 10^7$, the model reaches its minimum loss after going through 97\% of the training budget (141 data passes out of 147 scheduled), whereas for $D=6.4\times 10^6$ the corresponding number is 3\% (6 data passes out of 196 scheduled). 

\begin{figure}
    \centering
    \includegraphics[width=0.95\linewidth]{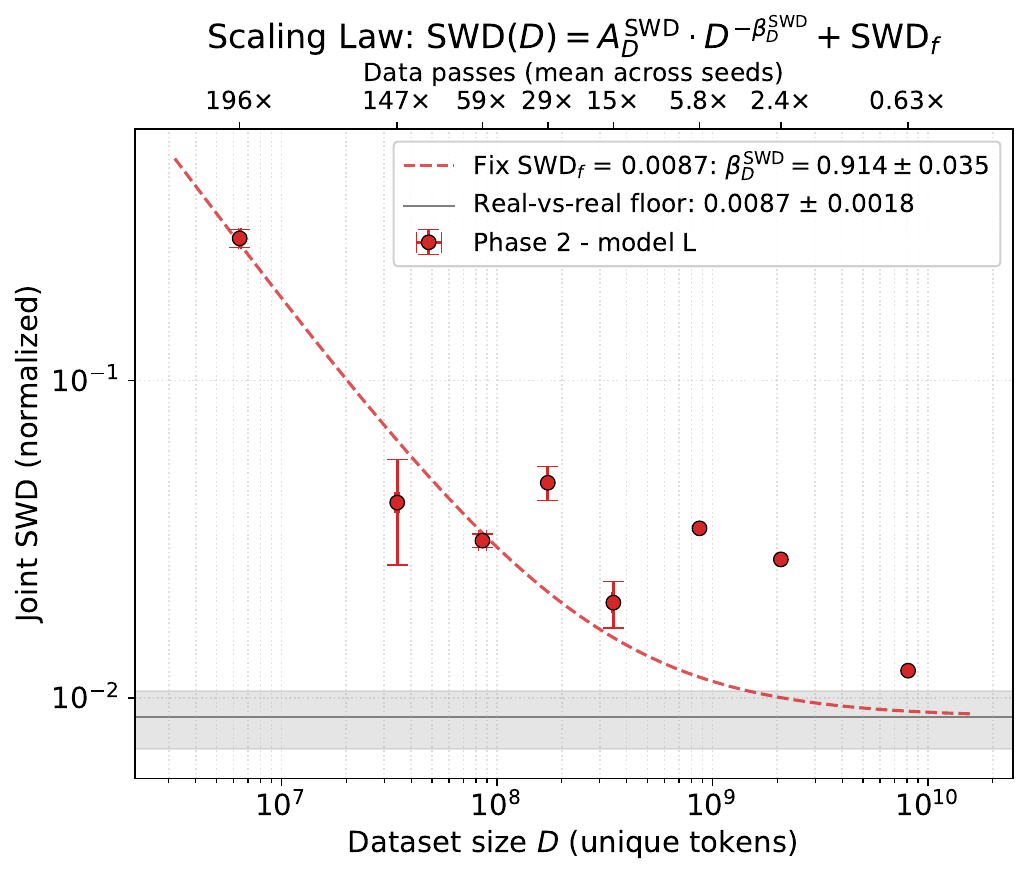}
    \caption{SWD fitted to a scaling law as a function of dataset size. The numbers correspond to the number of files used. The fit is performed with the SWD floor fixed to the real-vs-real noise floor. For $D\leq 3.5\times 10^8$, the trainings were repeated three times with different seeds in order to mitigate any stochastic effects arising from the choice of files. Horizontal error bars (barely visible) show the standard deviation of $D$ across seeds. The vertical per-point uncertainties were also taken from the variance over seeds, where available (the three largest $D$ were single-seed runs).}
    \label{fig:phase2c_swd}
\end{figure}

Figure \ref{fig:phase2c_swd} shows the SWD performance for the phase 2 runs. With the measured real-vs-real noise floor SWD$_f=0.0087\pm 0.0019$ as per phase 1, the power law
\begin{equation}
    \mathrm{SWD}(D) = A_D^{\mathrm{SWD}} D^{-\beta_D^{\mathrm{SWD}}}+\mathrm{SWD}_f
\label{eq:swd_d}
\end{equation}
has parameters 
\begin{equation*}
    A_N^{\mathrm{SWD}}=(4.3 \pm 3.2)\times 10^5, \qquad \beta_N^{\mathrm{SWD}} = 0.914\pm 0.035.
\end{equation*}

\subsection{Phase 3: compute}
\label{sec:results_phase_3}
\begin{figure}
    \centering
    \includegraphics[width=0.95\linewidth]{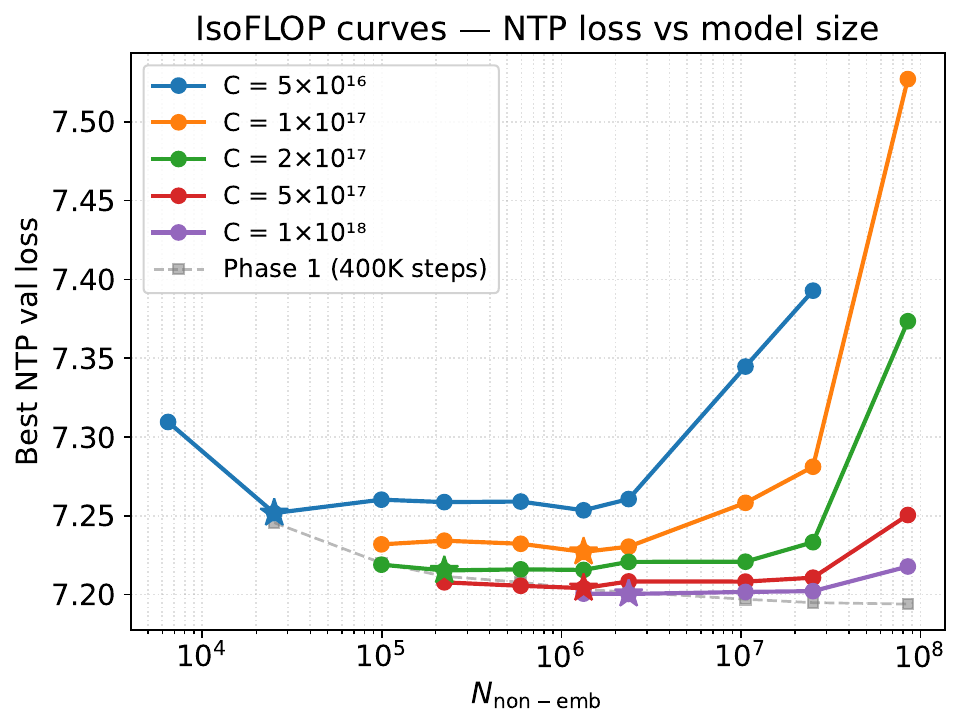}
    \caption{IsoFLOP curves showing the best validation NTP loss as a function of non-embedding parameters $N_\mathrm{non-emb}$. Note that the fit was performed with $N_\mathrm{incl}$ as described in the text, but $N_\mathrm{non-emb}$ is shown here to be comparable to the plots of phase 1. The discrete lowest NTP loss per compute budget is marked with a star.}
    \label{fig:isoflop_curves_Nincl}
\end{figure}

\begin{figure}
    \centering
    \includegraphics[width=0.95\linewidth]{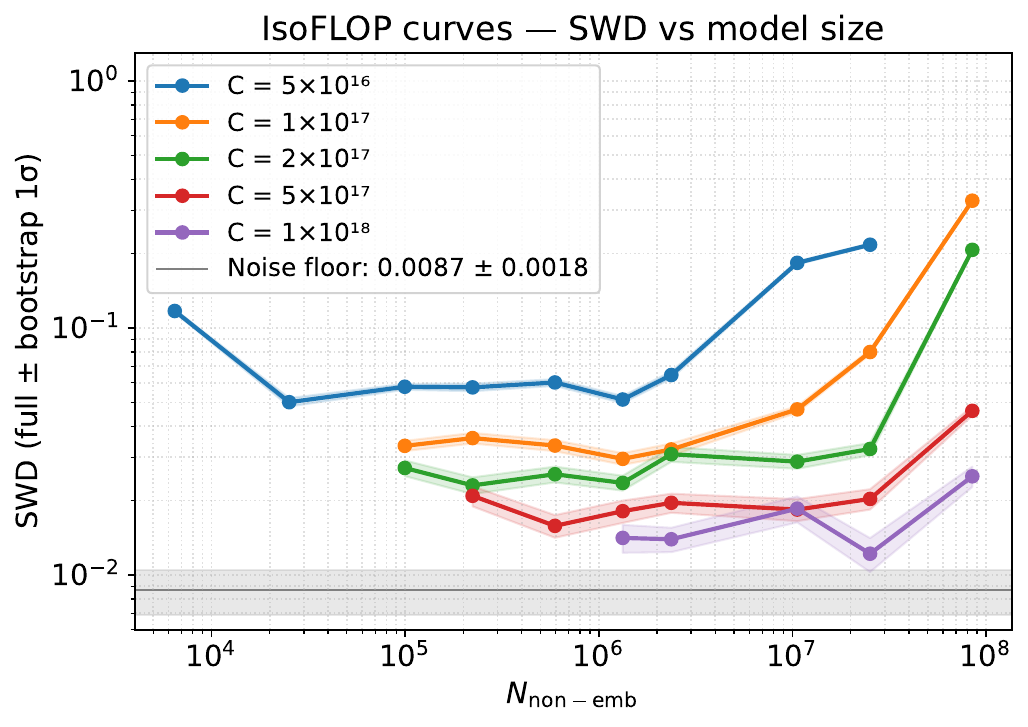}
    \caption{IsoFLOP curves showing the SWD as a function of model size. Note that the fit was performed with $N_\mathrm{incl}$ as described in the text, but $N_\mathrm{non-emb}$ is shown here to be comparable to the plots of phase 1. The error bands show the bootstrap standard deviation. The noise floor and its uncertainty band is the real-vs-real SWD$_f$ from phase 1.}
    \label{fig:isoflop_swd}
\end{figure}

Fig. \ref{fig:isoflop_curves_Nincl} shows the best NTP validation loss for each combination of model sizes and isoFLOPs as per Tab.~\ref{tab:isoflops}, and Fig.~\ref{fig:isoflop_swd} shows the corresponding SWD.

In a typical scaling regime, outside of any double descent behavior, it is expected that these isoFLOP curves show a U-shape: first model performance improves as a function of the number of parameters, until the model becomes overparameterized and begins to overfit the training data, leading to increasing loss values.
This is what has been observed both for language data~\cite{hoffmann2022trainingcomputeoptimallargelanguage} and jet classification~\cite{Vigl:2026ppx}.
In contrast, we do not see much of the left flank of the expected U-shape, the performance does not degrade significantly with low numbers of parameters.

Nevertheless, it is possible to use a parabolic fit to extract the optimal model size $N^*$ for each compute budget $C$. A fit to the power law
\begin{equation}
    N^*(C) \propto C^a
\end{equation}
results in 
\begin{equation*}
    a = 0.92\pm 0.18,
\end{equation*}
regardless of whether the fit is based on the NTP loss or the SWD results. Although the two methods mostly agree on the best model size for each compute budget tested, the optimal point is not well resolved due to the flat U curves and the absence of a clear left flank. 

\section{Discussion}
\label{sec:discussion}
The three phases show that the scaling of validation loss and SWD with model size $N$, dataset size $D$ and compute budget $C$ is compatible with power laws. The cleanest power law behavior is observed for the scaling with $N$, while the scaling with $D$ and $C$ are considerably less clear. 

A non-trivial observation is that the next token prediction loss $L_{\mathrm{NTP}}$ and high-level SWD metric behave similarly. Within the range probed, improved next token prediction on the jet constituents directly leads to improved modeling of physically relevant jet-level observables. 

An additional striking result of the study is how small the relative learnable window is. 
The small size of the learnable window is especially noteworthy when compared to prior results on text data which first observed the scaling phenomena. 
Table~\ref{tab:learnable_window} shows a comparison of various metrics related to the learnable window between this work and~\cite{hoffmann2022trainingcomputeoptimallargelanguage}. Since the codebook sizes are similar, $\log |V|$ is also similar. However, the $L_{\infty}$ of this work is considerably larger than that of~\cite{hoffmann2022trainingcomputeoptimallargelanguage}, leading to a much smaller learnable window $W$. The effective perplexity $e^{L_{\infty}}$ is a measure for how many effectively equiprobable tokens the asymptotic predictor has to choose from. In our study this is 1330, compared to the 5 of~\cite{hoffmann2022trainingcomputeoptimallargelanguage}. We also note that although our study spans a larger range in $N$ and $D$ than~\cite{hoffmann2022trainingcomputeoptimallargelanguage}, the exploited fraction of the window, $\Delta L_{\mathrm{NTP}}/W$, is considerably smaller.

\begin{table}
\centering
\begin{tabular}{lcc}
\toprule
Diagnostic & This work & Hoffmann et al.~\cite{hoffmann2022trainingcomputeoptimallargelanguage} \\
\midrule
$|V|$ & 32,768 & 32,000 \\
$\log |V|$ [nats] & 10.397 & 10.374 \\
$L_{\infty}$ (or $E$) [nats] & 7.193 & 1.69 \\
$W = \log |V| - L_{\infty}$ [nats] & 3.204 & 8.684 \\
$L_{\infty} / \log |V|$ & 0.692 & 0.163 \\
$\exp(L_{\infty})$ [effective tokens] & 1330 & 5.42 \\
\midrule
$\Delta L_{\mathrm{NTP},N} / W$ & $ 1.6\%$ & --- \\
$\Delta L_{\mathrm{NTP},D} / W$ & $ 11.9\%$ & --- \\
$\Delta L_{\mathrm{NTP},D\geq 3.5\times 10^7} / W$ & $ 1.3\%$ & --- \\
$\Delta L / W$ (full grid, $N$ + $D$) & --- & $ 15\%$ \\
\bottomrule
\end{tabular}
\caption{Comparison of learnable-window diagnostics between this work and the Chinchilla scaling-law study of Ref.~\cite{hoffmann2022trainingcomputeoptimallargelanguage}. Despite similar vocabulary sizes, jet next token prediction exhibits a substantially smaller learnable window and much larger effective perplexity.}
\label{tab:learnable_window}
\end{table}

As the dynamic range of the NTP loss is only a small fraction of $W$, these measurements sit in a low signal-to-noise regime, where the returns are diminishing and the stochasticity from run to run and the ensuing fits approaches the order of magnitude of the reducible loss. Although the fit in phase 1 is tight, it operates over just 1.6\% of $W$. Even the smallest model (Pico) captures 98.4\% of the learnable window. For $D=3.5\times 10^7$ and $D=8.6\times 10^7$ in phase 2, the seed to seed variability spans almost the entire range of $\Delta L_{\mathrm{NTP}}$ for all data points with $D \geq 3.5\times 10^7$ files. In phase 3, the nominal message of $a=0.92$ in $N^*(C)\propto C^a$ is that all extra compute should be allocated to model size. However, the flatness of the parabolas and the absence of the left flank (for most of the curves) means that the fit must be taken with a grain of salt.

A possible interpretation of this behavior is related to the structure of the AOJ dataset itself. In phase 2, even the extreme $D=6.4\times 10^6$ run, which saturates after having gone through only 3\% of its training budget, reaches 88\% of $W$. Recent work on the theoretical interpretation of neural scaling laws has suggested that the observed power-law behavior may reflect structural properties of the underlying data distribution rather than merely optimization effects. In particular, Ref.~\cite{bahri2024explaining} distinguishes between regimes where performance gains arise from resolving progressively finer structure in the underlying data distribution and regimes where the remaining uncertainty is dominated by intrinsic stochasticity. Our observations are qualitatively consistent with the latter picture. The dominant contribution to the AOJ dataset, approximately 99.5\%, consists of QCD jets. 
The remaining 0.5\% of the dataset is expected to contain jets with more substructure, like jets originating from vector bosons, Higgs bosons or top quarks. 
In such a situation, increasing the model or dataset size may primarily reduce fluctuations around an already well-resolved distribution, naturally leading to a compressed learnable window, weak dependence on dataset size, and rather flat isoFLOP minima. 

Though QCD jets contain a rich internal structure, as encoded in parton shower programs~\cite{Sjostrand:2006za,Bellm:2015jjp,Gleisberg:2008ta} and confirmed in various measurements \cite{Kogler:2018hem, ALICE:2021aqk, ATLAS:2020bbn, CMS:2017qlm, CMS:2023lpp}, the underlying dynamics are quantum mechanical, which may make autoregressive prediction of individual constituents highly stochastic.
This may be further exacerbated by the imposed $p_T$ ordering of the particles for the autoregressive task being in conflict with the underlying shower dynamics. 
A parton shower is often modeled with a tree structure \cite{Dreyer:2018nbf, Andreassen:2018apy}, with the initial parton seeding the shower and then spawning new partons according to some splitting function. 
The $p_T$ ordering we have imposed for autoregressive generation does not respect this causal structure; higher $p_T$ constituents from one branch of the tree may provide very little useful context to predict the next-highest $p_T$ constituent from a different part of the shower tree. 
In future work it would be interesting to compare the scaling behavior of alternative ordering schemes better suited for jet data. 

Another interesting direction for future work concerns the optimal codebook size. Physics data has an inherent resolution due to the constraints of the measurement. If the codebook size is too large, this might introduce a degeneracy in predicting the next token, which leads to increased stochasticity. A follow-up study would investigate not only the approximate resolution of the particle flow jet constituents, but also the capabilities of the VQ-VAE to align its codebook accordingly. The methodology from such a study could potentially be transferred to other domains, allowing for determining the optimal codebook size (or informing the choice to use a non-tokenized model) depending on the intrinsic resolution of the data.

The comparatively weak scaling gains observed in this study
also need to be put in context with the successful scaling results which have recently been observed on supervised jet classification tasks \cite{ATLAS:2026zdb, Vigl:2026ppx}. 
In those studies classification performance was seen to improve across many orders of magnitude in data and model size, and performance was not yet observed to saturate. 
Supervised classification learns the likelihood ratio between the signal and background classes, while NTP learns the likelihood of the data.
The AOJ data used for NTP is dominated by QCD jets, which form the background class for the supervised trainings. 
Continued scaling on supervised classification may indicate that the additional information being learned by the model pertains to the details of the signal class rather than the background class; ie the continued improvements come from better learning the numerator rather than the denominator of the likelihood ratio. 
Alternatively, it could indicate that the bulk of the QCD structure is learnable relatively easily, leading to the plateau observed in this study, but that there is additional behavior in its tails which requires scale to learn. 
This tail behavior could form a small fraction of the total QCD distribution and thus offer a minor contribution to the NTP loss, but nevertheless be highly relevant for classification tasks and benefit from scale. 

It should also be noted that the classification studies mentioned here use a richer set of low-level inputs, including track-level information, while our study is restricted to the basic kinematic features available in AOJ. This suggests another possible explanation for the different scaling behavior: the taggers may benefit from scaling because they have access to a larger and more informative feature space.
Future work studying the scaling behavior on different jet types as well as the effect of adding additional low-level features could clarify this point.

\section{Conclusion}
\label{sec:conclusion}

This work presented the first study of empirical scaling laws for foundation models applied to particle physics collision data. In three phases (varying dataset size, varying number of parameters at fixed dataset size, and varying number of parameters at fixed compute) the scaling behavior of \oja{} trained on Aspen Open Jets data was tested using two metrics: the loss function of the model and the sliced Wasserstein distance (SWD) as a statistical measure of the quality of the generated jets.

This is a critical addition as the SWD possesses a statistical lower limit that is indeed approached by the larger models. 
Since we observe a monotonic relation between the SWD and validation loss, this also allows comparing asymptotic behaviour in both metrics. We find that the asymptotic loss, the asymptotic SWD, and the statistical limit of the SWD metric are broadly consistent within uncertainties in the phase 1 scaling of model size, where a clear scaling law can be observed. 

While distinct logarithmic scaling behavior is observed as a function of model size, the behavior as a function of dataset size and compute was more difficult to interpret. 
We introduce the concept of a learnable window, the fraction of the next token prediction loss taken up by the learnable objective rather than stochastic effects.
We find that next token prediction on jet constituents has a significantly smaller learnable window than language tasks, suggesting a possible explanation for the comparatively weak scaling gains and rapid saturation observed in this study. 

Overall, our results demonstrate that further exploration and understanding is needed to achieve scalable unsupervised pre-training on particle physics data. This is especially relevant given the large volumes of collider data potentially available for (unsupervised) pre-training. 

Pre-training strategies which have demonstrated major successes on language data may need domain-specific adaptations when applied to other forms of scientific data.  
Particle physics data can therefore act as a laboratory to study the scaling behavior of large foundation models under conditions where large data volumes with full control over content (collider, simulation, fast simulation, AI surrogate simulation) exist.

\acknowledgments
The authors thank Alexander M\"uck and Ian Pang for contributing to discussions during the project, and for valuable comments on the manuscript.

TG, AH and GK acknowledge funding via the SciFM consortium (05D25GU4) funded by the German Federal Ministry of Research, Technology, and Space (BMFTR) in the ErUM-Data action plan, as well as support from the Deutsche Forschungsgemeinschaft (DFG, German Research Foundation) under the German Excellence Initiative – EXC 2121 Quantum Universe – 390833306. This research was supported in part through the Maxwell computational resources operated at Deutsches Elektronen-Synchrotron DESY, Hamburg, Germany. MK and HRG acknowledge funding via the SciFM consortium (05D25PA5) funded by the German Federal Ministry of Research, Technology, and Space (BMFTR) in the ErUM-Data action plan, as well as support from the German Research Foundation (DFG) under grant 396021762 -- TRR 257: Particle Physics Phenomenology after the Higgs Discovery.
O.A. is supported byFermi Forward Discovery Group, LLC under Contract No. 89243024CSC000002 with the U.S. Department of Energy, Office of Science, Office of High Energy Physics. This report is available under Fermilab open technical publications under FERMILAB-PUB-26-0345-PPD. D.S.\, and D.A.F.\, are supported by the DOE under Award Number DOE-SC0010008. This research used resources of the National Energy Research Scientific Computing Center (NERSC), a Department of Energy User Facility using NERSC award HEP-ERCAP0036658.

\bibliographystyle{apsrev4-2}
\bibliography{references}

\appendix
\section{Methodology}
\label{app:methodology}

\subsection{Search space and hyperparameters}
\label{app:search_space}
\begin{figure}[!hb]
    \centering
    \includegraphics[width=0.85\linewidth]{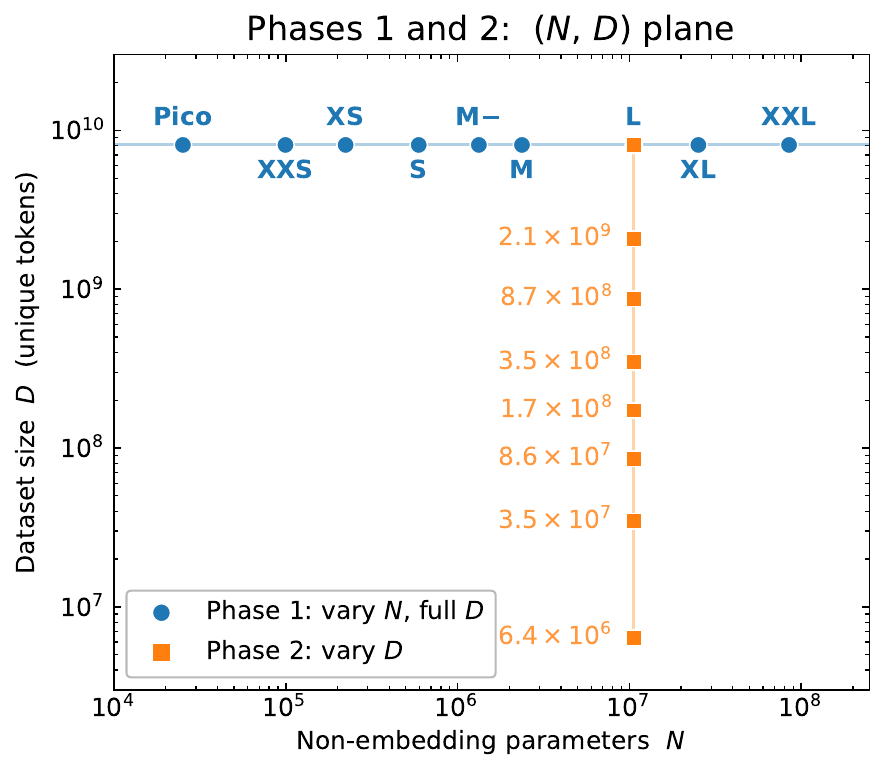}
    \caption{Search space in the $N$ - $D$ plane for phase 1 and 2, showing the model sizes $N$ and dataset sizes $D$. All models in phase 1 have access to the full dataset. For phase 2, the model L is chosen since it is large enough to not become a bottleneck, but still reasonable given the available computational resources.}
    \label{fig:methodology_ND_space}
\end{figure}

\begin{figure}
    \centering
    \includegraphics[width=0.85\linewidth]{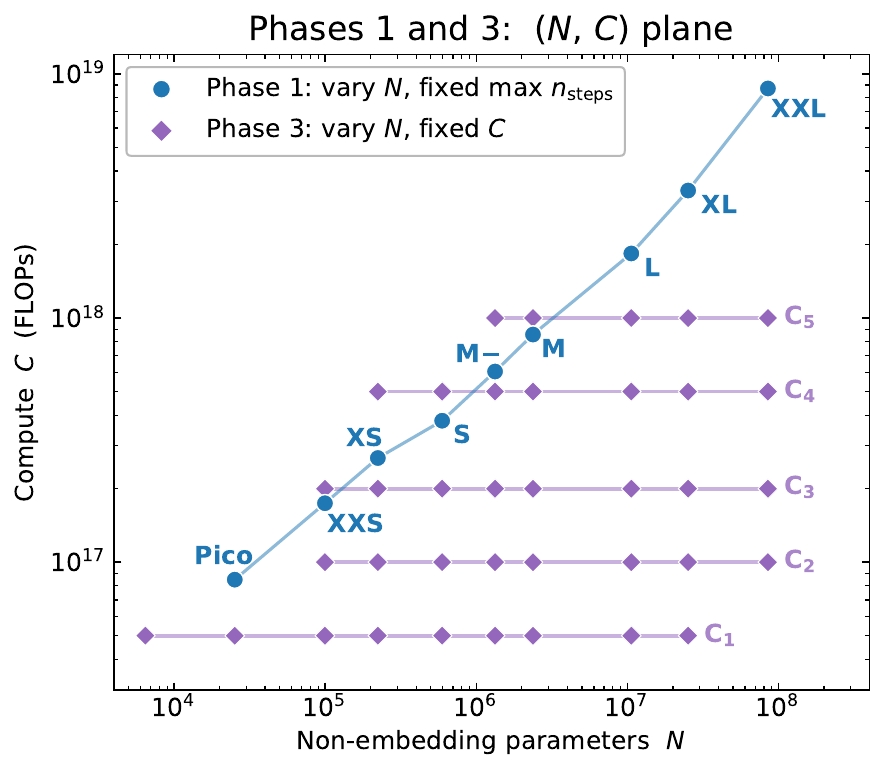}
    \caption{Search space in the $N$ - $C$ plane, showing the combinations of model sizes $N$ and compute budgets $C$.}
    \label{fig:methodology_NC_space}
\end{figure}

Figure~\ref{fig:methodology_ND_space} shows the search space for phase 1 and 2, and Fig.~\ref{fig:methodology_NC_space} shows the search space for phase 1 and 3. The model transformer architectures are specified in Tab.~\ref{tab:model_sizes}, the dataset sizes in Tab.~\ref{tab:phase2_dataset_sizes}, and the number of gradient steps for each compute budget and model combination in Tab.~\ref{tab:isoflops}. 

Phase 1 and 2 models are trained for 400k steps, with 200 validation cycles. All models are trained with a cosine annealing schedule with LR$_{min}=0$, where $T_{max}$ is set to the full 400k steps in phase 1 and 2 (except for $D=6.4\times 10^6$ in phase 2, where $T_{max}=100k$, as discussed in Sec.~\ref{sec:method}), and to the full training schedule of Tab.~\ref{tab:isoflops} for phase 3. All training runs across all phases use a batch size of 256. Ranger~\cite{Ranger}, which combines RAdam~\cite{Liu2020On} with 
Lookahead~\cite{NEURIPS2019_90fd4f88}, is chosen as optimizer. The learning rates are decided using a learning rate sweep, with the final ones listed in Tab.~\ref{tab:learning_rates}.

\begin{table}
\centering
\begin{tabular}{lcccccc}
\toprule
Size & $d_{\mathrm{emb}}$ & $H$ & $L$ & $N_{\mathrm{non-emb}}$ & $N_{\mathrm{incl}}$ \\
\midrule
Sub-Pico & 16  & 2  & 2  & $6.46 \times 10^{3}$ & $5.31 \times 10^{5}$ \\
Pico     & 32  & 2  & 2  & $2.52 \times 10^{4}$ & $1.07 \times 10^{6}$ \\
XXS      & 64  & 2  & 2  & $9.96 \times 10^{4}$ & $2.20 \times 10^{6}$ \\
XS       & 96  & 4  & 2  & $2.23 \times 10^{5}$ & $3.37 \times 10^{6}$ \\
S        & 128 & 4  & 3  & $5.94 \times 10^{5}$ & $4.79 \times 10^{6}$ \\
M-       & 192 & 8  & 3  & $1.33 \times 10^{6}$ & $7.62 \times 10^{6}$ \\
M        & 256 & 8  & 3  & $2.37 \times 10^{6}$ & $1.08 \times 10^{7}$ \\
L        & 384 & 8  & 6  & $1.06 \times 10^{7}$ & $2.32 \times 10^{7}$ \\
XL       & 512 & 8  & 8  & $2.52 \times 10^{7}$ & $4.20 \times 10^{7}$ \\
XXL      & 768 & 12 & 12 & $8.50 \times 10^{7}$ & $1.10 \times 10^{8}$ \\
\bottomrule
\end{tabular}
\caption{Model size configurations and parameter counts for the different models, specifying the embedding dimension $d_{\mathrm{emb}}$, the number of attention heads $H$, the number of GPT blocks $L$, and the resulting parameter count with ($N_{\mathrm{incl}}$) and without ($N_{\mathrm{non-emb}}$) the final unembedding layer.}
\label{tab:model_sizes}
\end{table}

\begin{table}
\centering
\begin{tabular}{ccc}
\toprule
$D$ (tokens)  & Seed runs & Data passes \\
\midrule
$6.4 \times 10^{6}$ &     3 & $196$ \\
$3.5 \times 10^{7}$ &     3 & $147$ \\
$8.6 \times 10^{7}$ &     3 &  $59$ \\
$1.7 \times 10^{8}$ &     3 &  $29$ \\
$3.5 \times 10^{8}$ &     3 &  $15$ \\
$8.7 \times 10^{8}$ &     1 & $5.8$ \\
$2.1 \times 10^{9}$ &     1 & $2.4$ \\
$8.1 \times 10^{9}$ &     1 & $0.63$ \\
\bottomrule
\end{tabular}
\caption{Dataset sizes with corresponding number of data passes in the phase 2 study.}
\label{tab:phase2_dataset_sizes}
\end{table}

\begin{table}
\centering
\begin{tabular}{lccccc}
\toprule
 & \multicolumn{5}{c}{Compute budget $C$ (FLOPs)} \\
\cmidrule(lr){2-6}
Model & $5\times10^{16}$ & $1\times10^{17}$ & $2\times10^{17}$ & $5\times10^{17}$ & $1\times10^{18}$ \\
\midrule
Sub-Pico & 475K & --- & --- & --- & --- \\
Pico     & 235K & --- & --- & --- & --- \\
XXS      & 115K & 230K & 459K & --- & --- \\
XS       & 75K  & 150K & 300K & 749K$\dagger$ & --- \\
S        & 53K  & 105K & 211K & 527K & --- \\
M$^{-}$  & 33K  & 66K  & 132K & 331K & 662K$\dagger$ \\
M        & 23K  & 47K  & 93K  & 234K & 467K \\
L        & 11K  & 22K  & 44K  & 109K & 218K \\
XL       & 6K   & 12K  & 24K  & 60K  & 120K \\
XXL      & ---  & 5K   & 9K   & 23K  & 46K \\
\midrule
\# models & 9 & 8 & 8 & 7 & 5 \\
\bottomrule
\end{tabular}
\caption{Number of gradient steps for each model and compute budget combination in the isoFLOP study of phase 3. Two runs ($\dagger$) have a data repetition of 4-17\%, the others have no data repetition. Some combinations of model size and compute budget have been excluded from the study according to the criteria outlined in the main text.}
\label{tab:isoflops}
\end{table}

\begin{table}
\centering
\begin{tabular}{lccc}
\toprule
Model & Learning rate \\
\midrule
Pico & $1 \times 10^{-2}$ \\
XXS  & $5 \times 10^{-3}$ \\
XS   & $5 \times 10^{-3}$ \\
S    & $3 \times 10^{-3}$ \\
M$^{-}$ & $3 \times 10^{-3}$ \\
M    & $3 \times 10^{-3}$ \\
L    & $1 \times 10^{-3}$ \\
XL   & $1 \times 10^{-3}$ \\
XXL  & $3 \times 10^{-4}$ \\
\bottomrule
\end{tabular}
\caption{Learning rates used across all phases for the different model sizes.}
\label{tab:learning_rates}
\end{table}

\subsection{Sliced Wasserstein distance}
\label{app:swd}
The Sliced Wasserstein distance~\cite{Bonneel2015} is a compute-efficient replacement of the Wasserstein-1 distance in higher dimensions, that reduces the $d$-dimensional problem to a collection of $1d$ problems via the Radon~\cite{kolouri2019generalizedslicedwassersteindistances} transform. 

In general, the 1-Wasserstein distance is defined as 
\begin{equation}
W_{n,m}=\int_{\mathbb{R}} |F_n(u)-G_m(u)|\,du
\end{equation}
where \(F_n,G_m\) are empirical CDFs. For equal sample sizes $m=n$, it is equivalent to compute the average distances over the ordered samples $x_i$ and $x'_i$, thus the 1D Wasserstein can be written as
\begin{equation}
W_n=\frac{1}{n}\sum_{i=1}^{n}\left|\underline{x}_i-\underline{x}'_i\right|.
\end{equation}

The $\mathrm {SWD}$ is then given by
\begin{equation}
\mathrm {SWD} = \displaystyle\frac{1}{L} \sum_{\theta\in\Omega_K}\left(\frac{1}{n}\sum_{i=1}^{n} \left| \underline{x}_{i}^{\theta}-\underline{x}_{i}'^{\theta}\right|\right)
\end{equation}
where $\Omega_L$ is a set of $L$ randomly selected directions on the unit sphere $\Omega=\{\theta\in\mathbb{R}^d\setminus||\theta||=1\}$, and $\{\underline{x}_i^\theta\}_{i=1}^n=\{\theta^T x_i\}_{i=1}^n$ are the sorted data points projected on the direction $\theta$ (see \cite{Grossi:2024axb}). 

In practice, the integral over the sphere in the Radon transform is replaced by a Monte Carlo average over $L$ random directions. After standardizing the five distributions ($p_\mathrm{T}, m, \tau_{21}, \tau_{32}, n$) by subtracting the mean and dividing by the standard deviation of the real sample, $L=500$ random unit vectors are drawn. Each of these are used to project the $5d$ space onto a single dimension, where the Wasserstein-1 distance is calculated. The final SWD is obtained by averaging over all 500 1D Wasserstein distances.

\section{The NTP loss}
\label{app:loss}
In an autoregressive transformer trained with NTP, the model the probability of a full sequence $x=(x_1,\dots,x_T)$ is given by
\begin{equation}
q_{\theta}(x)=\prod_{t=1}^{T} q_{\theta}(x_t \mid x_{<t}),
\end{equation}
where $x_{<t}=(x_1,\dots,x_{t-1})$ denotes the previous tokens. During training, the target is to minimize the average next token cross-entropy loss,
\begin{equation}
\mathcal{L}_{\mathrm{NTP}}
=
-\mathbb{E}_{x\sim p}
\left[
\frac{1}{T}\sum_{t=1}^{T}
\log q_\theta(x_t\mid x_{<t})
\right],
\end{equation}
which is equivalent to the negative log-likelihood of the full sequence distribution induced by the model.

Let $p(x)$ be the true data distribution and $q_\theta(x)$ be the model distribution parameterized by $\theta$. Then the cross-entropy of $p$ relative to $q_\theta$ is
\begin{equation}
H(p, q_\theta) = - \mathbb{E}_{x \sim p} \big[ \log q_\theta(x) \big].
\end{equation}

This is exactly the negative log-likelihood (NLL) that the model minimizes during training. Furthermore, the entropy of the true distribution is
\begin{equation}
H(p) = - \mathbb{E}_{x \sim p} \big[ \log p(x) \big].
\end{equation}

This represents the irreducible loss $L_{\infty}$, the minimum achievable cross-entropy.

Moreover, the KL divergence from $p$ to $q_\theta$ can be written as
\begin{equation}
D_{\mathrm{KL}}(p \,\|\, q_\theta) = H(p, q_\theta) - H(p) 
\end{equation}

Since the training objective corresponds to minimizing the empirical negative log-likelihood, it estimates the cross-entropy $H(p, q_\theta)$. Thus, the KL divergence measures the excess cross-entropy above the entropy of the data distribution.

\section{Normalization: Nats per Token}

Since different sequences can have different lengths $T$, it is standard to normalize
the loss by the number of elements in the sequence:
\begin{equation}
\mathcal{L}_{\text{per-token}}(x; \theta)
= - \frac{1}{T} \sum_{t=1}^{T} \log q_\theta(x_t \mid x_{<t}).
\end{equation}

Averaging over the validation dataset $\mathcal{D}_{\text{val}}$ gives the expected
\emph{cross-entropy per token},
\begin{equation}
L_{\text{val}} = \mathbb{E}_{x \sim \mathcal{D}_{\text{val}}}[\mathcal{L}_{\text{per-token}}(x; \theta)].
\end{equation}

This quantity has units of \emph{nats per token} (if natural logarithms are used) or
\emph{bits per token} (if base-2 logarithms are used).

\vfill
\pagebreak

\end{document}